# Numerical-analytical algorithm for constructing the envelope of the projectile trajectories in midair

## Peter S Chudinov

Perm State Agricultural Academy, Perm, 614990, 25-th October st., 10, Russia
E-mail: chupet@mail.ru

## Abstract

A classic problem of the motion of a point mass (projectile) thrown at an angle to the horizon is reviewed. The air drag force is taken into account with the drag factor assumed to be constant. An analytic approach is mainly used for the investigation. Simple analytical formulas are used for the constructing the envelope of the family of the point mass trajectories. The equation of envelope is applied for the determination of the maximum range of flight. The motion of a baseball is presented as an example.

## 1. Introduction

The problem of the motion of a point mass thrown at an angle to the horizon is a constituent of many introductory courses on physics. This task arouses interest of authors as before [1– 3]. With zero air drag force, the analytic solution is well known. The trajectory of the point mass is a parabola. With air drag taken into account, a finding the main variables of the problem is reduced to quadratures.  Appropriate integrals are not taken in finite form. The problem, to all appearances, does not have the exact analytic solution, and therefore in most cases is solved numerically [4 – 9]. Analytic approaches to the solution of the problem are not sufficiently advanced. Meanwhile, analytic solutions are very convenient for a straightforward adaptation to applied problems and are especially useful for a qualitative analysis. Comparatively simple approximate analytical formulas to study the motion of the point mass in a medium with quadratic drag force have been obtained using such an approach [10 – 13] . These formulas make it possible to carry out a complete qualitative analysis without using the numeric integration of point mass motion differential equations.

This paper considers the application of formulas [10 – 13] for the constructing the envelope of the family of the point mass trajectories. The family of trajectories is formed when throwing a point mass with one and same initial velocity, but with different angles of throwing. The problem of constructing the envelope is solved by means of simple analytical formulas. The use of numerical integration differential equations of motion of a point mass is minimal. The equation of the envelope is used for finding of the maximum range of flight of point mass in the case when the spot of incidence is above or below than the spot of throwing. In the cited examples it is shown that the obtained formulas ensure a high accuracy of the determination of the maximum range. The mistake of analytical calculation of range does not exceed 1%.

## 2. Equations of point mass motion and analytical formulas for basic parameters

The problem of the motion of a point mass in air, with a number of conventional assumptions, in case of the drag force proportional to the square of the velocity, $R = mgkV^2$, boils down to a numerical integration of the differential system [14]

$$\frac{dV}{dt} = -g\sin\theta - gkV^2, \quad \frac{d\theta}{dt} = -\frac{g\cos\theta}{V}, \quad \frac{dx}{dt} = V\cos\theta, \quad \frac{dy}{dt} = V\sin\theta. \quad (1)$$

Here $V$ is the velocity of the point mass, $\theta$ is the angle between the tangent to the trajectory of the point mass and the horizontal, $g$ is the acceleration due to gravity, $m$ is the mass of the particle, $x$, $y$ are the Cartesian coordinates of the point mass, $k = \frac{\rho_a c_d S}{2mg} = const$ is the proportionality factor, $\rho_a$ is the air density, $c_d$ is the drag factor for a sphere, and $S$ is the cross-section area of the object.

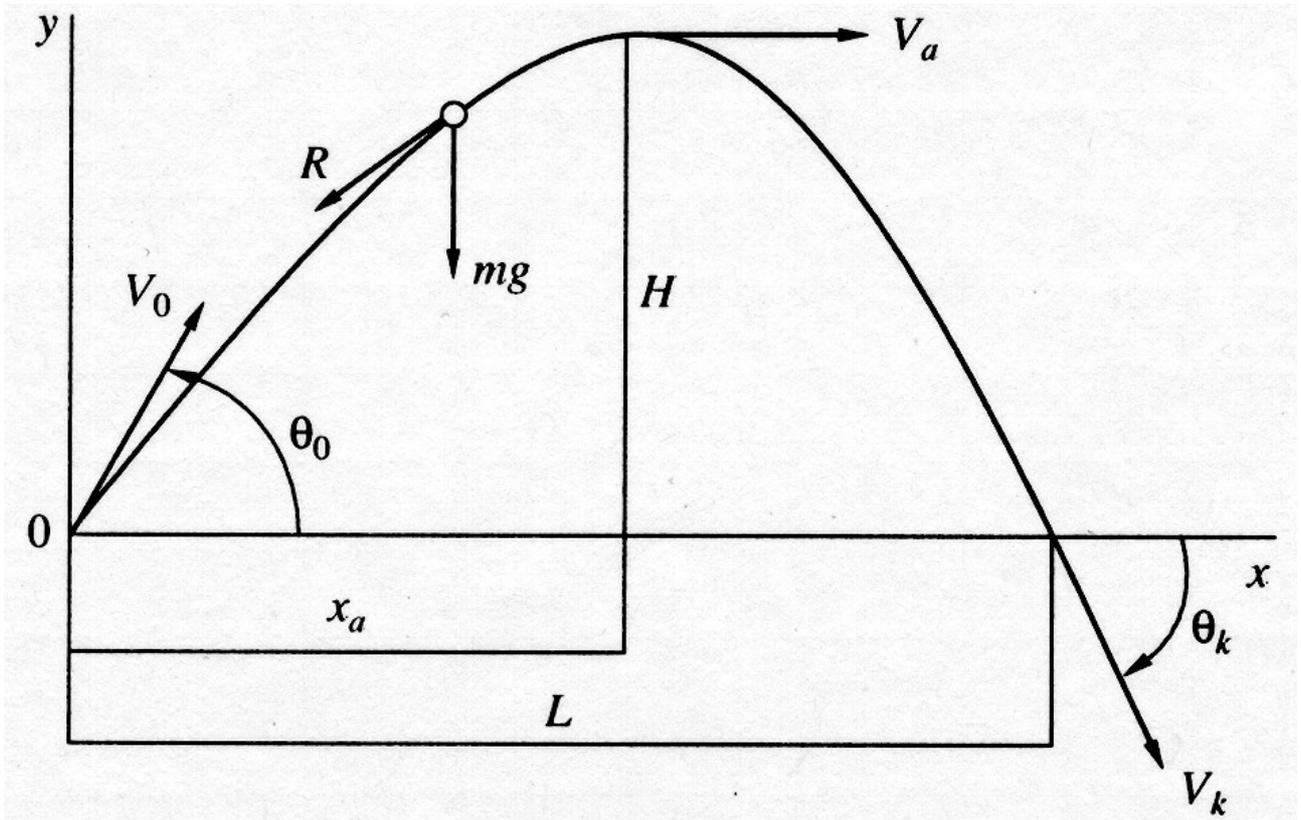

**Figure 1.** Basic motion parameters.

Comparatively simple approximate analytical formulas for the eight basic parameters of point mass motion are obtained in [10 – 13]. Let us set out the relationships required for the maximum ascent height $H$, motion time $T$, velocity at the trajectory apex $V_a$, flight range $L$, and the abscissa of the trajectory apex $x_a$ (Figure 1):

$$H=\frac{V_0^2\sin^2\theta_0}{g(2+kV_0^2\sin\theta_0)} \quad , \quad T=2\sqrt{\frac{2H}{g}} \quad , \quad V_a=\frac{V_0\cos\theta_0}{\sqrt{1+kV_0^2(\sin\theta_0+\cos^2\theta_0\ln(\tan(\frac{\theta_0}{2}+\frac{\pi}{4})))}} \quad ,$$

$$L=V_a T \quad , \quad x_a=\sqrt{LH\cot\theta_0} \quad . \tag{2}$$

With zero drag ($k = 0$), formulas (2) go over into the respective formulas of the point mass parabolic motion theory. All motion characteristics described by relationships (2) are functions of $V_0$, $\theta_0$, initial conditions of throwing. Relationships (2), in turn, make it possible to obtain simple analytical formulas for six basic functional relationships of the problem which are $y(x)$, $y(t)$, $x(t)$, $y(\theta)$, $x(\theta)$, $t(\theta)$ [11].

One of the most important aspects of the problem is the determination of an optimum angle of throwing of a point mass which provides the maximum range. The equation for the optimum angle of throwing α in the case when the points of incidence and throwing are on the same horizontal is obtained in [12]:

$$\tan^2\alpha+\frac{p\sin\alpha}{4+4p\sin\alpha}=\frac{1+p\lambda}{1+p(\sin\alpha+\lambda\cos^2\alpha)} \quad . \tag{3}$$

Here $p=kV_0^2$, $\lambda(\alpha)=\ln(\tan(\frac{\alpha}{2}+\frac{\pi}{4}))$. We use formulas (2) - (3) for the constructing the envelope.

## 3. The equation of the envelope in midair

In the case of no drag the trajectory of a point mass is a parabola. For the different angles of throwing under one and same initial velocity the projectile trajectories form a family of parabolas. The maximum range and the maximum height for limiting parabolas are given by formulas

$$L_{max}=\frac{V_0^2}{g} \quad , \quad H_{max}=\frac{V_0^2}{2g} \quad . \tag{4}$$

The envelope of this family is also a parabola, the equation of which is usually written as [15]

$$y(x)=\frac{V_0^2}{2g}-\frac{g}{2V_0^2}x^2 \quad . \tag{5}$$

Using (4), we will convert the equation (5) as

$$y(x)=\frac{H_{max}(L_{max}^2-x^2)}{L_{max}^2} \quad . \tag{6}$$

We will set up an analytical formula similar to (6) for the envelope of the point mass trajectories taking into account the air drag force.

Taking into account the formula (6), we will derive an equation of the envelope as

$$y(x) = \frac{H_{max}(L_{max}^2 - x^2)}{L_{max}^2 - ax^2} \quad . \tag{7}$$

Such structure of the equation takes into account the fact that the envelope has a maximum under $x = 0$. Besides, function (7) under $a > 0$ has a vertical asymptote, as well as any point mass trajectory accounting the resistance of air. In formula (7) $H_{max}$ is the maximum height, reached by the point mass when throwing with initial conditions, $V_0$, $\theta_0 = 90°$; $L_{max}$ - the maximum range, reached when throwing a point mass with the initial velocity $V_0$ under some optimum angle $\theta_0 = \alpha$. In the parabolic theory an angle $\alpha$ is $45°$ under any initial velocity. Taking into account the resistance of air, an optimum angle of throwing $\alpha$ is less than $45°$ and depends on the value of parameter $p = kV_0^2$. Parameter $H_{max}$ with the preceding notation is defined by formula [8]

$$H_{max} = \frac{1}{2gk} \ln(1 + kV_0^2) \quad . \tag{8}$$

A choice of a positive factor $a$ in the formula (7) is sufficiently free. However it must satisfy the condition $a = 0$ in the absence of resistance ($k = 0$). We shall find this coefficient from the following considerations.

Investigation [11] shows that, while taking into account air resistance, the trajectory of a point mass is well approximated by the function

$$y(x) = \frac{Hx(L-x)}{x_a^2 + (L - 2x_a)x} \quad . \tag{9}$$

Here *x, y* are the Cartesian coordinates of the point mass; parameters *H, L, $x_a$* are shown in Fig. 1.

Thus, for the generation of the equation of the maximum range trajectory it is required three parameters: *H, $L_{max}$, $x_a$*. We will calculate these parameters as follows. Under a given value of quantity $p = kV_0^2$ we will find the root $\alpha$ of equation (3). An angle $\alpha$ ensures the maximum range of the flight. By integrating numerically system (1) with the initial conditions $V_0$, $\alpha$, we obtain the values $H(\alpha)$, $L_{max} = L(\alpha)$, $x_a(\alpha)$ for the maximum range trajectory. We set the equal slopes the tangents to envelope (7) and to the maximum range trajectory

$$y(x) = \frac{H(\alpha) x(L_{max} - x)}{x_a^2(\alpha) + (L_{max} - 2x_a(\alpha))x}$$

in the spot of incidence $x = L_{max}$. It follows that parameter $a$ is defined by formulas

$$a = 1 - \frac{2H_{max}}{H(\alpha)} \left(1 - \frac{x_a(\alpha)}{L_{max}}\right)^2 \quad . \tag{10}$$

In the absence of air resistance parameter $a$ vanishes.

The equation of the envelope can be used for the determination of the maximum range if the spot of falling lies above or below the spot of throwing. Let the spot of falling be on a horizontal straight line defined by the equation $y = y_1 = const$. We will substitute a value $y_1$ in the equation (7) and solve it for *x*. We obtain the formula

$$x_{max} = L_{max} \sqrt{\frac{H_{max} - y_1}{H_{max} - ay_1}} . \qquad (11)$$

The correlation (11) allows us to find a maximum range under the given height of the spot of falling.

## 4. The results of the calculations

As an example we will consider the moving of a baseball with the resistance factor $k = 0.000548$ s$^2$/m$^2$ [6]. Other parameters of motion are given by values

$$g = 9.81 \text{ m/s}^2, \ V_0 = 50 \text{ m/s}, \ y_1 = \pm 20, \pm 40, \pm 60 \text{ m}.$$

Substituting values *k* and $V_0$ in the formula (8), we get $H_{max} = 80.26$ m. Hereinafter we solve an equation (3) at the value of non-dimensional parameter $p = kV_0^2 = 1.37$. The root of this equation gives the value of an optimum angle of throwing. This angle ensures the maximum range: $α = 40°$. By integrating the system of equations (1) with the initial conditions

$$V_0 = 50, \ θ_0 = 40°, \ x_0 = 0, \ y_0 = 0 ,$$

we find meanings

$$H(α) = 36.2 \text{ m}, \ L_{max} = L(α) = 133.6 \text{ m}, \ x_a(α) = 75.1 \text{ m}. \qquad (12)$$

According to the formula (10) the factor is $a = 0.149$. The graph of the envelope (7) is plotted in Figure 2 together with the family of trajectories. We note that family of trajectories is received by means of numerical integrating of the equations of motion of a point mass (1). A standard fourth-order Runge-Kutta method was used.

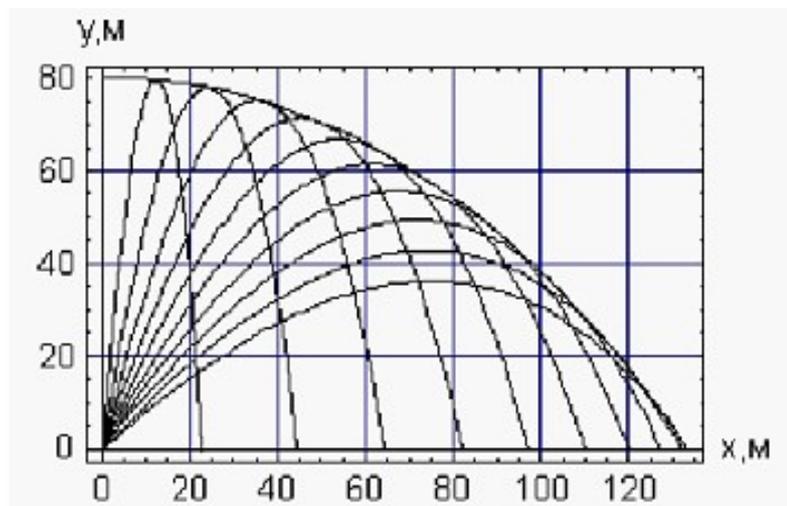

**Figure 2.** The family of projectile trajectories and the envelope of this family.

The results of calculations using the formula (11) are presented in table 1.

Table 1. Maximum range under different heights of the spot of the falling.

| $y_1$ (m) | Analytical value $x_{max}$, (m) | Numerical value $x_{max}$, (m) | Error ( % ) |
|---|---|---|---|
| 60 | 71.2 | 71.1 | 0.1 |
| 40 | 98.3 | 98.3 | 0. |
| 20 | 118.0 | 118.0 | 0. |
| 0 | 133.6 | 133.6 | 0. |
| -20 | 146.6 | 146.5 | 0.1 |
| -40 | 157.8 | 157.5 | 0.2 |
| -60 | 167.5 | 166.9 | 0.4 |

The second column of the table contains range values calculated analytically by the formula (11). The third column of the table contains range values from integration of the equations of motion (1). The fourth column presents an error of the calculation of the range in the percentage. Error does not exceed 1 %. Formula (11) gives almost the exact value of the maximum range in a wide range of height point drop ( 120 m ). The tabulated data show that formulas (2), (3), (7), (11) ensure sufficiently pinpoint accuracy of the calculation of parameters of motion.

Note that values of $H(\alpha)$, $L_{max} = L(\alpha)$, $x_a(\alpha)$ can be obtained by using the formulas (2), without numerical integration of the system (1). Substituting $V_0$ and $\alpha$ in the formulas (2), we have

$$H(\alpha) = 36.5 \text{ m}, \quad L_{max} = L(\alpha) = 132.4 \text{ m}, \quad x_a(\alpha) = 76.0 \text{ m}. \tag{13}$$

For these values $a = 0.2$. The graph of the envelope does nearly change (Figure 3). The right end of the graph shift along the $x$ axis is less than 1%.

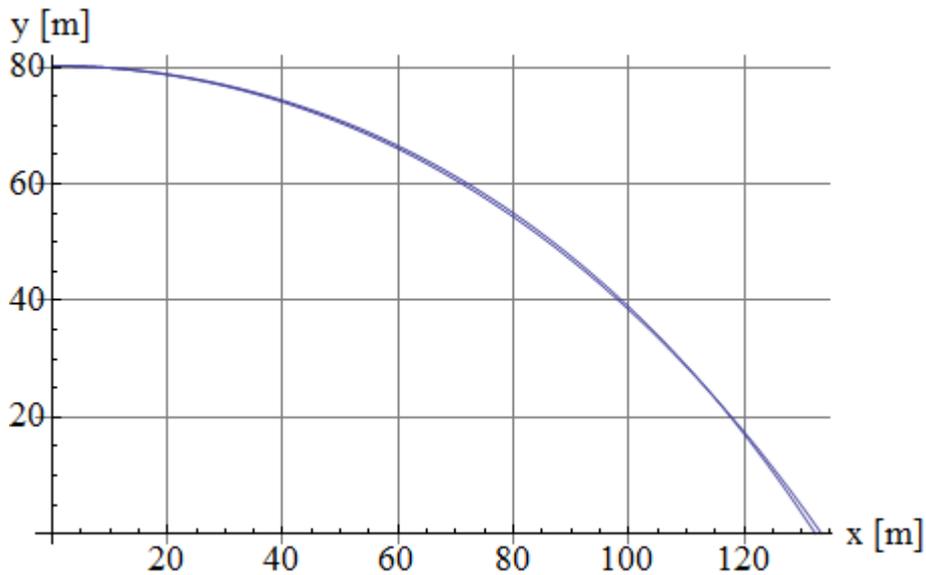

Figure 3. Graphs of envelopes were constructed with the help of values (12) and (13).

## 5. Conclusion

The proposed approach based on the use of analytic formulas makes it possible to significantly simplify a qualitative analysis of the motion of a point mass with the air drag taken into account. All basic parameters of motion, functional relationships and various problems of optimization are described by simple analytical formulas. Moreover, numerical values of the sought variables are determined with an acceptable accuracy. Thus, many formulas [10 – 13] and (2) – (11) make it possible to carry out complete analytical investigation of the motion of a point mass in a media with drag just as it is done for the case of no drag.